\documentclass[12pt]{article}

\newcommand{\be}{\begin{equation}}
\newcommand{\ee}{\end{equation}}
\newcommand{\bn}{\begin{eqnarray}}
\newcommand{\en}{\end{eqnarray}}

\newcommand{\uf}{{U(\varphi,\varphi^*)}}

\newcommand{\lmmcs }{{\cal L}_{M+MCS}}
\newcommand{\lmcs }{{\cal L}_{MCS}}
\newcommand{\lsd}{{\cal L}_{SD}}
\newcommand{\lm}{{\cal L}_{{\rm master}}}
\newcommand{\lmsd}{{\cal L}_{M+SD}}
\newcommand{\jo}{J_{\nu}^{(0)}}

\newcommand{\nn}{\nonumber}
\newcommand{\no}{\noindent}

\def\bea{\begin{eqnarray}}
\def\eea{\end{eqnarray}}

\newcommand{\beq}{\begin{equation}}
\newcommand{\eeq}{\end{equation}}

\begin{document}

\title{\textbf{On the coupling of the self-dual field to dynamical U(1) Matter and
its dual theory }}
\author{D. Dalmazi \\
\textit{{UNESP - Campus de Guaratinguet\'a - DFQ} }\\
\textit{{Av. Dr. Ariberto Pereira da Cunha, 333} }\\
\textit{{CEP 12516-410 - Guaratinguet\'a - SP - Brazil.} }\\
\textsf{E-mail: dalmazi@feg.unesp.br }}
\date{\today}
\maketitle

\begin{abstract}
We consider an arbitrary $U(1)$ charged matter non-minimally
coupled to the self-dual field in $d=2+1$. The coupling includes a
linear and a rather general quadratic term in the self-dual field.
By using both a Lagragian gauge embedding and a master action
approaches we derive the dual  Maxwell Chern-Simons type model and
show the classical equivalence between the two theories. At
quantum level the master action approach in general requires the
addition of an awkward extra term to the Maxwell Chern-Simons type
theory. Only in the case of a linear coupling in the self-dual
field the extra term can be dropped and we are able to establish
the quantum equivalence of gauge invariant correlation functions
in both theories.

\textit{{PACS-No.:} 11.15.-q, 11.10.Kk, 11.10.Gh, 11.10.Ef }
\end{abstract}



\newpage

\section{Introduction}


It is very common in physics that due to technical difficulties we
are not able to probe, even at theoretical level, all physical
aspects of a given theory. In those cases it is important to have
an equivalent ( dual ) description of the same theory. A classical
example is the dual description of the massive Thirring model in
$d=1+1$ by means of the massive Sine-Gordon model
\cite{Coleman,Mandelstam} (see also \cite{Abdalla} and references
therein). The strong coupling expansion in one model corresponds
to the usual weak coupling expansion in the other one and
vice-versa. Some nonperturbative features like confinement can
also be revealed with the help of duality as in \cite{SW}. In this
work we are interested in the specific case of dual descriptions
of a spin one massive particle in $d=2+1$ (see \cite{TPN,DJ}).
This type of duality naturally appears in  the bosonization
program  in $d=2+1$ where fermionic $U(1)$ current correlators in
the massive Thirring model can be equally described by a bosonic
spin one massive field which, for infinitely massive fermions, may
be either of self-dual (SD) or Maxwell Chern-Simons (MCS) type
\cite{FS,Kondo,Banerjee1,Barci,ddh2,BottaHelayel}.

By means of a master action approach it was shown in \cite{DJ}
that the gauge invariant sector of the MCS theory is on shell
equivalent to the SD theory with the dual map
$f_{\mu}\leftrightarrow
\frac{\epsilon_{\mu\nu\alpha}\partial^{\nu}A^{\alpha}}m$. The
equivalence holds beyond the classical level and includes gauge
invariant correlation functions in the quantum theory as shown in
\cite{Banerjee2}. Although the equivalence between those  free
theories is interesting in its own, the most powerful applications
of duality occur in interacting theories. The case of the
non-abelian version of a possible SD/MCS duality is much less
trivial and it is still under investigation \cite{nonabelian}.
Another possibility to build an interacting theory is to couple
the vector fields in SD and MCS theories to $U(1)$ charged matter
fields. Along this direction it has been shown in \cite{GMS} that
the SD field minimally coupled to $U(1)$ charged massive fermions
is dual equivalent to a MCS gauge field coupled to matter through
a Pauli term. Besides, a Thirring term must be added to assure the
equivalence also in the fermionic sector. The demonstration of
\cite{GMS}, which is based on a master action, guarantees the
equivalence of the equations of motion as well as of the partition
functions of the dual theories. More recently, in \cite{Anacleto2}
both cases of $U(1)$ charged fermions and scalar fields minimally
coupled to the SD field have been considered through a Lagrangian
gauge embedding procedure. In particular, the results of
\cite{GMS} have been reproduced at classical level pointing out
that the dual map must include now a contribution from the matter
fields as we will see later in this work. The case of scalar
fields minimally coupled to the SD field leads to a MCS theory in
a field dependent medium, a highly nonlinear theory. Once again,
the minimal coupling is replaced by a Pauli term in the dual MCS
type theory and a current-current term must be introduced to
guarantee equivalence in the scalar sector.  Similar applications
of the Lagrangian gauge embedding procedure to produce gauge
theories dual to non-gauge theories have been worked out, for
instance in \cite{Bazeia,Menezes}. Only the classical equivalence
has been considered in \cite{Anacleto2,Bazeia,Menezes}. In those
cases the matter fields are always minimally coupled to the vector
field. Since those vector fields are not gauge fields there is no
reason {\it a priori} to use minimal coupling. In other words, it
is not clear why should one couple those fields to a conserved
current, as already remarked in \cite{DJ}. The aim of this work is
to clarify this issue in the literature by assuming a rather
general coupling between the self-dual field and an arbitrary
$U(1)$ charged matter. In section 2 we present a simple proof of
on shell equivalence between the corresponding SD and MCS theories
non-minimally coupled to dynamical $U(1)$ matter. The proof is
independent of the details of the matter theory.  In section 3 we
analyze the equivalence of gauge invariant correlation functions
in the quantum theory and draw some conclusions in section 4.

\section{On shell equivalence }

\subsection{ Lagrangian gauge embedding }

We start from the Lagrangian of the self-dual field coupled to some $U(1)$ charged
matter fields. Due to the mass term in the SD theory
we have no gauge invariance and
therefore we are not constrained to use the minimal coupling condition. The coupling
will be rather general and will contain a linear and a quadratic term in the self-dual
field such that the global $U(1)$ symmetry of the matter theory
is preserved, namely

\be \lmsd \, = \, \lsd + \frac{U(\varphi , \varphi^*) }2
f^{\nu}f_{\nu} -e f^{\nu}J_{\nu}^{(0)} + {\cal L}^{U(1)}_{{\rm
Matter}}\quad .  \label{lmsd} \ee

\no Where we define:

\bea {\cal L}_{SD} \, &=& \, \frac{m^2}2 f^{\mu}f_{\mu} - \frac
m2\epsilon_{\alpha\beta\gamma}
f^{\alpha}\partial^{\beta}f^{\gamma} \quad , \label{sd} \\ {\cal
L}^{U(1)}_{{\rm Matter}} \, &=& {\cal L}_{{\rm Kinetic }} +
V(\varphi , \varphi^*) \label{lmatter} \eea The generic notation
$(\varphi , \varphi^*)$ represents the matter fields which may be
either fermions, scalars or even complex vector fields. The
function of the matter fields $U(\varphi , \varphi^*)$ is positive
and global $U(1)$ invariant, but otherwise arbitrary. Likewise,
$V(\varphi , \varphi^*)$ is also assumed to be an arbitrary and
global $U(1)$ invariant function which may include a mass term
while ${\cal L}_{{\rm Kinetic }}$ represents the purely kinetic
terms. The current $\jo$ is the $U(1)$ Noether current which comes
from the free matter theory ${\cal L}_{{\rm Kinetic }}$.
Minimizing the action we obtain the equations of motion : \be
K_{\nu} \, \equiv \, \frac{\delta S}{\delta f^{\nu}}  \, = \,
\left(\mu^2 g_{\nu\alpha} + m
\epsilon_{\nu\alpha\gamma}\partial^{\gamma}\right)f^{\alpha} - e
\jo \, = 0 \, ,\label{euler} \ee

\be \int d^3x\left( \delta {\cal L}^{U(1)}_{{\rm Matter}} - e
f^{\nu}\delta\jo + \frac{\delta\mu^2}2 f^{\nu}f_{\nu} \right) = 0
\quad . \label{eqmatter} \ee Where we use the notation: \be \mu^2
\, = \, m^2 + U(\varphi , \varphi^*) \label{mu2} \ee

\no Dropping the matter fields, it is known [5,6] that $\lsd $ is dual
to the Lagrangian
\be {\cal L}_{MCS} \, = \, \frac m2\epsilon_{\alpha\beta\gamma}
A^{\alpha}\partial^{\beta}A^{\gamma} - \frac 14 F_{\mu\nu}(A)F^{\mu\nu}(A).
\label{mcs} \ee

\no Therefore, one might ask what is the Lagrangian dual to $\lmsd
$. The basic idea is to add extra terms to (\ref{lmsd}) such that
the new theory is invariant under the gauge transformation $\delta
f_{\nu}=
\partial_{\nu}\alpha \, ; \, \delta \varphi = 0 $ while keeping its
equivalence to the old theory. Adapting the procedure of
\cite{Anacleto2} to our case leads to : \be \lmmcs \, = \, \lmsd -
\frac 1{2\mu^2} K^{\nu}K_{\nu} \label{kk} \ee It is a simple
exercise to check that $\lmmcs $ is invariant under the
aforementioned gauge transformations.  Using $\, \delta
K_{\nu}=\mu^2\delta f_{\nu} + \delta\mu^2 f_{\nu} -e \delta \jo +
\epsilon_{\nu\alpha\gamma}\partial^{\gamma}\delta f^{\alpha}\, $,
after an integration by parts and a trivial cancellation, we can
write down the equations of motion of $\lmmcs $ in the form :

\be \epsilon^{\nu\gamma\alpha}\partial_{\alpha}\left(\frac{K_{\gamma}}{\mu^2}\right) =
0 \, \, , \label{3} \ee

\be \int d^3x\left\lbrack \delta {\cal L}^{U(1)}_{{\rm Matter}} - e
\left(f^{\nu}-\frac{K^{\nu}}{\mu^2}\right)\delta\jo + \frac{\delta\mu^2}2
\left(f^{\nu}-\frac{K^{\nu}}{\mu^2}\right)^2 \right\rbrack = 0 \quad . \label{4} \ee

\no From the definition of $K^{\nu}$ in  (\ref{euler}) we have the
identity \be -m
\epsilon_{\nu\alpha\gamma}\partial^{\gamma}\left(\frac{K^{\alpha}}{\mu^2}\right)
= \left( \mu^2 g_{\nu\alpha} + m
\epsilon_{\nu\alpha\gamma}\partial^{\gamma}\right)
\left(f^{\alpha}-\frac{K^{\alpha }}{\mu^2}\right)-e\jo
\label{identity} \ee Thus, (\ref{3}) and (\ref{4}) are equivalent
to \be \left(\mu^2 g_{\nu\alpha} + m
\epsilon_{\nu\alpha\gamma}\partial^{\gamma}\right){\tilde
f}^{\alpha} - e \jo \, = 0 \, ,\label{5} \ee \be \int d^3x\left(
\delta {\cal L}^{U(1)}_{{\rm Matter}} - e {\tilde
f}^{\nu}\delta\jo + \frac{\delta\mu^2}2 {\tilde f}^{\nu}{\tilde
f}_{\nu} \right) = 0 \quad . \label{6} \ee which have the same
form of the equations of motion of $\lmsd$, see (\ref{euler}) and
(\ref{eqmatter}), with $f_{\nu}$ being replaced by : \be {\tilde
f}_{\nu} \, = \, f_{\nu}-\frac{K_{\nu}}{\mu^2}\, =\, -\frac
m{\mu^2} \epsilon_{\nu\alpha\gamma}\partial^{\gamma}f^{\alpha} +
\frac e{\mu^2}\jo \label{7} \ee

\no The equations (\ref{5}) and (\ref{6}) are gauge invariant as a
consequence of the gauge invariance of ${\tilde f}_{\nu}$. Solving
(\ref{5}) for ${\tilde f}_{\nu}= {\tilde f}_{\nu}(\varphi,
\varphi^*)$ and plugging back in (\ref{6}) will furnish equations
of motion involving only matter fields which are, of course, the
same equations obtained by solving (\ref{euler}) for $f_{\nu}=
f_{\nu}(\varphi, \varphi^*)$ and substituting back in
(\ref{eqmatter}). That proves the on shell equivalence of $\lmmcs$
and $\lmsd$ in the matter sector. Concerning the equations of
motion for $f_{\alpha}$, notice that this is a gauge field in
$\lmmcs$ and its equation of motion will depend in general on some
arbitrary space-time function. We can make this arbitrariness
explicit by noticing that the most general solution to (\ref{3})
is: \be K_{\nu} \, = \, \mu^2
\partial_{\nu} \alpha \, = \, \mu^2\left(g_{\nu\beta}+ m
\epsilon_{\nu\beta\gamma}\right) f^{\beta} - e \jo \quad ,
\label{solution} \ee where $\alpha (x_{\mu})$ is an  arbitrary
space-time function. By choosing  $\alpha $  a constant function
we recover equation (\ref{euler}) and the on shell equivalence
between $\lmsd$ and $\lmmcs$ is fully demonstrated for arbitrary
coupling $U\left(\varphi,\varphi^*\right)$ and  generic $U(1)$
matter fields with arbitrary potential
$V\left(\varphi,\varphi^*\right)$. In particular, no requirement
of minimal coupling was necessary for the demonstration. For
future use we end up this subsection writing down the full MCS
type Lagrangian dual to $\lmsd$. From (\ref{kk}) we have : \bea
\lmmcs = &-&\frac{m^2}{4\mu^2}F_{\alpha\beta}(A)F^{\alpha\beta}(A)
+ \frac m2
\epsilon_{\alpha\beta\gamma}A^{\alpha}\partial^{\beta}A^{\gamma} \nn \\
&-& \frac{m \, e}{\mu^2}\jo
\epsilon^{\nu\alpha\beta}\partial_{\alpha}A_{\beta} -
\frac{e^2}{2\mu^2}\jo J^{\nu \, (0)} + {\cal L}_{{\rm
matter}}^{U(1)} \, . \label{lmmcs} \eea As usual, we have renamed
$f_{\nu}\to A_{\nu}$. The above highly nonlinear Lagrangian is
gauge invariant ($\delta A_{\nu}=\partial_{\nu}\alpha $) and
contains a Pauli and a Thirring like term as in the simpler case
of minimal coupling  \cite{Anacleto2}.

\subsection{Master Action}

The classical equivalence between the free theories $\lsd$ and
$\lmcs$ was proved in \cite{DJ} using a master action approach.
This was later generalized in \cite{GMS} for the coupling
(minimal) with fermions. Here we suggest a new master action which
interpolates between $\lmsd $ and  $\lmmcs $ for arbitrary $U(1)$
matter. This approach is specially useful when comparing
correlation functions in the quantum theory
\cite{Banerjee2,Bazeia2}.

First of all, notice that the first three terms in (\ref{lmmcs}) can be combined such
that \be \lmmcs = -\frac{\mu^2}2\left(\frac{m\epsilon_{\nu\alpha\beta}\partial^{\beta}
A^{\alpha}-e \jo }{\mu^2}\right)^2  + \frac m2
\epsilon_{\alpha\beta\gamma}A^{\alpha}\partial^{\beta}A^{\gamma} + {\cal L}_{{\rm
matter}}^{U(1)} \, , \label{8} \ee This formula suggests the gauge invariant master
Lagrangian: \bea \lm \, &=& \, \frac{\mu^2}2 f^{\nu}f_{\nu} +
f^{\nu}\left(m\epsilon_{\nu\alpha\beta}\partial^{\beta}
A^{\alpha}-e \jo \right) \nn \\
& + & \frac m2 \epsilon_{\alpha\beta\gamma}A^{\alpha}\partial^{\beta}A^{\gamma} +
{\cal L}_{{\rm matter}}^{U(1)} \, \label{9} \eea where $\mu^2$ is defined as in
(\ref{mu2}) and $f_{\nu}$ is so far an extra non-dynamical auxiliary field. It is easy
to show that the equations of motion from $\lm $ are equivalent to the equations of
motion that we get from $\lmmcs $ or $\lmsd $. Indeed, minimizing $\lm $ we have:

\be \epsilon^{\gamma\beta\alpha}\partial_{\beta}(A_{\alpha}-f_{\alpha}) \, = \, 0
\label{10} \ee \be \mu^2 f_{\nu} \, =\, e \jo - m
\epsilon_{\nu\gamma\delta}\partial^{\delta}A^{\gamma} \label{11}\ee \be \int
d^3x\left( \delta {\cal L}^{U(1)}_{{\rm Matter}} - e f^{\nu}\delta\jo +
\frac{\delta\mu^2}2 f^{\nu}f_{\nu} \right) = 0 \quad . \label{12} \ee

\no From (\ref{10}) we have $A_{\nu}=f_{\nu}+\partial_{\nu}\alpha $ with arbitrary
$\alpha $. Plugging back in (\ref{11}) and (\ref{12}) we recover exactly the equations
of motion of $\lmsd $. Alternatively, solving (\ref{11}) for $f_{\nu}$ and returning
in (\ref{12}) we obtain the equations of motion of $\lmmcs $ in the matter sector,
while returning in (\ref{10}) we get exactly the equations of motion for the gauge
potential coming from $\lmmcs $, namely, \be
\epsilon^{\alpha\nu\beta}\partial^{\beta}\left( A_{\nu} - \frac{e\jo}{\mu^2} + \frac
m{\mu^2}\epsilon_{\nu\gamma\delta}\partial^{\delta}A^{\gamma}\right) \, = \, 0 \, ,
\nn \ee That completes the proof of on shell equivalence of $\lm $ to $\lmmcs$ and
$\lmsd$. Our master action works for arbitrary $U(1)$ matter  with no restrictions on
$V\left(\varphi,\varphi^*\right)$ and $U\left(\varphi,\varphi^*\right)$. The minimal
coupling has not played any role so far. However, there is one point  which deserves
to be mentioned. Namely, in the cases of minimal coupling analyzed in the literature
\cite{GMS,Anacleto2} and also for the free SD theory, the equation of motion for the
self-dual field, see (\ref{euler}), leads to $
\partial^{\nu} f_{\nu}=0\, $. This is no longer true for the
arbitrary coupling $\, U(\varphi,\varphi^*) \, $. For instance,
for scalar fields, where $J_{\nu}^{(0)}= \imath
\left(\phi\partial_{\nu}\phi^* - \phi^*\partial_{\nu}\phi
\right)$, we have instead :

\be
\partial_{\nu}\left\lbrace\left\lbrack m^2 + U(\phi,\phi^*)- 2 e^2 \phi^*\phi \right\rbrack
f^{\nu}\right\rbrace \, = \, 0 \quad . \label{df} \ee

\no Only in the case of minimal coupling, which corresponds to
$U(\phi,\phi^*) = 2 e^2 \phi^* \phi $,
 we obtain the simple relation $\partial_{\nu}f^{\nu}=0 $. This is quite surprising
because, as we stressed before, $\lmsd $ is not a gauge theory and we should not expect
any special results in the case of minimal coupling. Even from the point of view of
the dual  $MCS$ type theory this is not clear since this theory is much more
complicated for
$U(\phi,\phi^*) = 2 e^2 \phi^* \phi $ than it is for
$U(\phi,\phi^*) = 0$. One interesting point concerning the equation $\partial_{\nu}f^{\nu}=0 $
is the counting of degrees of freedom. Since, both free theories $\lsd $ and $\lmcs $
represent one polarization state of a massive spin one field and $\lsd $ is not a
gauge theory like $\lmcs $, it is necessary to have such equation to balance the
counting of degrees of freedom. It plays the same role in the SD theory
of a gauge condition in the MCS
one. Regarding our interacting theories, it is clear that the relation (\ref{df}) reduces the
number of degrees of freedom of $f_{\nu}$ by the same amount whatever we choose for
$U(\phi,\phi^*)$ as far as $U(\phi,\phi^*) \ne 2 e^2 \phi^* \phi - m^2 $. \footnote{We
are not interested in that special case in this work because we would have a gauge
theory governed by the Chern-Simons term instead of the self-dual theory.} Therefore,
to the best we know, there is no mandatory reason
to have the relation $\partial_{\nu}f^{\nu}=0 $ in the interacting theory.



\section{ Quantum equivalence }

In order to examine under which conditions the $\lmsd $ / $\lmmcs $
 duality holds at quantum level
we introduce the partition function :

\be {\cal Z} \, = \, \int  {\cal D}\varphi {\cal D}\varphi^* {\cal D}f^{\nu} {\cal
D}A^{\nu} \, e^{\, \imath\int d^3x\left\lbrack \lm + \frac{\lambda}2
(\partial^{\nu}A_{\nu})^2 \right\rbrack } \quad . \label{zmaster} \ee

\no Since $\lm $ is a gauge theory  we have added a gauge fixing term. The gauge field
$\, A_{\nu}\, $ can be easily integrated over. Using momentum space notation we
have: \bea & &\int  {\cal D}A^{\nu}e^{\frac {\imath}2\int \frac{d^3k}{(2\pi
)^3}\left\lbrack {\tilde A}^{\mu}(k){\cal O}_{\mu\nu}{\tilde A}^{\nu}(-k) + {\tilde
A}^{\mu}(k)T_{\mu}
(-k) + {\tilde A}^{\mu}(k)T_{\mu}(k) \right\rbrack }\nn \\
& = & C\, \exp -\frac {\imath}2\int \frac{d^3k}{(2\pi )^3}\, T_{\alpha }(k)\, ( {\cal
O}^{-1})^{\alpha\beta}T_{\beta}(-k)  \quad . \label{identity2} \eea

\no Where $\, C\, $ is a  numerical factor and \footnote{We use the same notation
, without tilde, for the Fourier components $f_{\nu}(k)$  and the space-time
components $f_{\nu}(x)$  to avoid confusion with the notation for the dual field (\ref{7}).}
\bea {\cal O}_{\alpha\beta}(k) \, &=& \, \imath \, m E_{\alpha\beta} + \lambda
k_\alpha k_\beta
\label{o}\\
{\cal O}^{-1}_{\alpha\beta}(k) \, &=& \, \frac {k_\alpha k_\beta}{\lambda \, k^4}
 + \imath  \frac{E_{\alpha\beta}}{m \, k^2}
\label{o-1} \\
T_{\mu}(-k) \, &=& \, -\imath m \, E_{\mu\alpha} f^{\alpha}(-k) \label{tmu} \\
E_{\mu\nu} \, &=& \, \epsilon_{\mu\nu\alpha}k^{\alpha} \quad . \label{emunu} \eea \no
Usually, a gaussian  integral like (\ref{identity2}) takes us from a local field
theory in space-time to a nonlocal one. Remarkably, thanks to the identity $\,
E_{\mu\nu}({\cal O}^{-1})^{\nu\alpha} E_{\alpha\beta} =
(-\imath/m)E_{\mu\beta}\, $ we end up with a local theory which is exactly $\lmsd $:

\be {\cal Z} \, = \, C \, \int  {\cal D}\varphi {\cal D}\varphi^* {\cal D}f^{\nu} \,
e^{\imath\int d^3x\lmsd }\quad . \label{zmsd} \ee

\no On the other hand, starting from (\ref{zmaster}) we could have made the
translation $\, f_{\nu}\to f_{\nu} + \left(e\jo - m \epsilon_{\nu\alpha\beta}\partial
^{\beta}A^{\alpha}\right)/\mu^2 \, $ leading to the dual partition function: \be {\cal
Z} \, = \, \int  {\cal D}\varphi {\cal D}\varphi^* {\cal D}f^{\nu} {\cal D}A^{\nu} \,
e^{\imath\int d^3x\left\lbrack \lmmcs + \frac{\lambda}2 (\partial^{\nu}A_{\nu})^2 +
{\cal L}_{{\rm extra }} \right \rbrack } \quad , \label{zextra} \ee

\no where $\lmmcs $ is given in (\ref{lmmcs}) and \be {\cal
L}_{{\rm extra }} \, = \, \frac{m^2+\uf }2 f^{\nu}f_{\nu} \quad .
\label{lextra} \ee \no At classical level, ${\cal L}_{{\rm extra
}}$ can be dropped since its only role is to produce the equation
of motion $f_{\nu}=0$. This is in agreement with the classical
equivalence between $\, \lmsd \, $ and $\, \lmmcs \, $ proved in
the last section. At quantum level we must distinguish the general
case $\, \uf \ne 0  \, $ from the case where the self-dual field
appears only linearly coupled with the matter fields $\, \uf =0 \,
$. In the first case we have the free propagator $\langle
f_{\alpha}(k) f_{\beta}(-k)\rangle =\frac{g_{\alpha\beta}}{m^2} \,
$. The vertex $\frac 12 \uf f^{\nu}f_{\nu} \, $ will then generate
new self-interaction vertices for the matter fields after
integration over $f_{\nu } $. All those new vertices have a
divergent coefficient (cubic divergence) due to the bad
ultraviolet behavior of the $f_{\nu}$ free propagator. In fact,
the same problem can be seen from the point of view of $\, \lmsd
\, $. There we have the free propagator of the self-dual field:
\be \langle f_{\alpha}(k) f_{\beta}(-k)\rangle_{M+SD}
=\frac{g_{\alpha\beta}}{k^2 - m^2} +
\frac{k_{\alpha}k_{\beta}}{m^2(k^2-m^2)}
 -\imath \frac{E_{\alpha\beta}}{m(k^2-m^2)} \quad . \label{fpropagator}
\ee As $k\to\infty $ the above propagator behaves like a constant,
which is a problem already noticed in \cite{Girotti,Anacleto1}.
Consequently the vertex $\frac 12 \uf f^{\nu}f_{\nu} \, $ present
in $\lmsd $ will also lead to nonrenormalizable cubic divergences
in the matter sector. Therefore it is not surprising to have such
divergences in the dual MCS type theory. In fact, the bad
ultraviolet behavior of the self-dual propagator is  a source of
troubles even for $\uf = 0 $ where it leads to power counting
nonrenormalizable interactions like the Thirring and the Pauli
term. Hopefully, one can surmount the problems of the linear
coupling case $\uf = 0$ by introducing extra flavors in the matter
sector and using a $1/N_f $ expansion. At least in the case of
fermionic matter fields there are power counting arguments given
in \cite{GMS} ( see also \cite{Freire}) pointing to a $1/N_f $
renormalizability although, a more detailed analysis and the
extension to  other types of matter fields  is missing in the
literature.

For the more general coupling $\uf \ne 0$ the introduction of
extra flavors is not enough to avoid the infinities generated by
${\cal L}_{extra}$ and one might try a more dramatic change in the
original SD theory. It is natural to add second order derivative
terms ( see similar trick in \cite{Freire} ) in order to improve
the ultraviolet behavior of the $f_{\nu}$ propagator. On one hand,
if we introduce a gauge non-invariant term like $(\lambda
/2)(\partial_{\nu}f^{\nu})^2 $, although the ultraviolet behavior
is certainly improved, we end up with a dual nonlocal MCS type
theory. Basically we have $\mu^2\to \mu^2
-\lambda\,\partial_{\nu}\partial^{\nu} $ and therefore $\,
1/(\mu^2 -\lambda\,
\partial_{\nu}\partial^{\nu} ) \,$ will be the source of nonlocality in the MCS type
theory. On the other hand, we can avoid nonlocality in the dual MCS theory by adding a
gauge invariant extra term like $\lambda \, F_{\mu\nu}(f)F^{\mu\nu}(f) $ but in this
case the ultraviolet behavior of $f_{\nu}$ will not be improved. In summary, there
seems to be no simple solution to the infinities appearing for $\uf \ne 0 $.

At this point we simply abandon the general case and stick to $\uf =0$ to avoid the
cubic divergences. In this case the  integral over $f_{\nu}$ in  (\ref{zextra})
produces only a numerical factor in the partition function : \be {\cal Z} \, = \,
{\tilde C}\, \int  {\cal D}\varphi {\cal D}\varphi^* {\cal D}A^{\nu} \, e^{\imath\int
d^3x\left\lbrack \lmmcs + \frac{\lambda}2 (\partial^{\nu}A_{\nu})^2 \right \rbrack }
\quad , \label{zmmcs} \ee Thus, comparing with (\ref{zmsd}), we conclude that in the
special case of linear coupling  the classical equivalence between $\lmmcs $ and
$\lmsd $ goes through the partition function. Likewise, we could have introduced
sources for the matter fields in (\ref{zmaster}) and since no integration over the
matter fields has been done so far, it would be trivial to show that the normalized
matter field correlation functions must be equal in both theories:

\be \left\langle \varphi_{a_1}(x_1)\cdots \varphi_{a_n}(x_n)
\varphi^*_{b_1}(y_1) \cdots \varphi^*_{b_n}(y_n)
\right\rangle_{M+MCS}^{U=0}
 = \left\langle \varphi_{a_1}(x_1)\cdots \varphi_{a_n}(x_n)
\varphi^*_{b_1}(y_1) \cdots \varphi^*_{b_n}(y_n)
\right\rangle_{M+SD}^{U=0} \label{mcorrelators} \ee Regarding the
vector field correlation functions,   we introduce two types of
sources in (\ref{zmaster}), as in the free case \cite{Banerjee2},
and define the generating function:

\be {\cal Z}(j,{\tilde j}) \, = \, \int  {\cal D}\varphi {\cal D}\varphi^* {\cal
D}f^{\nu} {\cal D}A^{\nu} \, e^{\, \imath\int d^3x\left\lbrack \lm + \frac{\lambda}2
(\partial^{\nu}A_{\nu})^2 + j^{\alpha}f_{\alpha} + {\tilde j}^{\alpha} {\tilde
F}_{\alpha}(A) \right\rbrack } \quad . \label{gf} \ee

\no Where ${\tilde F}_{\nu} (A) =
-(1/m)\epsilon_{\nu\beta\gamma}\partial^{\gamma}A^{\beta } + e
\jo/m^2 $. After the translation $f_{\alpha}\to f_{\alpha} +
{\tilde j}_{\alpha}/m^2 $ and integration over $A_{\mu}$ we obtain
:

\be {\cal Z}_{M+SD}(j,{\tilde j}) \, = \, C \int  {\cal D}\varphi
{\cal D}\varphi^* {\cal D}f^{\nu} {\cal D}A^{\nu} \, e^{\,
\imath\int d^3x\left\lbrack \lmsd +
 f_{\alpha}(j^{\alpha} + {\tilde j}^{\alpha}) + \frac{{\tilde j}^{\alpha}{\tilde
 j}_{\alpha}}{2 m^2} + \frac{j^{\alpha}{\tilde
 j}_{\alpha}}{ m^2}
 \right\rbrack } \quad . \label{gfsd} \ee

\no On the other hand, starting from (\ref{gf}) and integrating
over $f_{\nu}$ we derive

\be {\cal Z}_{M+MCS}(j,{\tilde j}) \, = \, {\tilde C} \int  {\cal
D}\varphi {\cal D}\varphi^* {\cal D}f^{\nu} {\cal D}A^{\nu} \,
e^{\, \imath\int d^3x\left\lbrack \lmmcs +
 {\tilde F}_{\alpha}(A)(j^{\alpha} + {\tilde j}^{\alpha}) -  \frac{j^{\alpha}
 j_{\alpha}}{ 2m^2}+ \frac{\lambda}2
(\partial^{\nu}A_{\nu})^2
 \right\rbrack } \quad . \label{gfmcs} \ee

 \no From (\ref{gfsd}) and (\ref{gfmcs}) we can obtain the correlation
 functions.
  For
instance, for the non-connected two and four-point correlation
functions we have:

 \bea \left\langle f_{\alpha_1}(x_1)
f_{\alpha_2}(x_2)\right\rangle_{M+SD}^{U=0}   &=&  \left\langle
{\tilde F}_{\alpha_1}(x_1){\tilde
F}_{\alpha_2}(x_2)\right\rangle_{M+MCS}^{U=0} -
i\delta_{\alpha_1\alpha_2}\frac{\delta (x_1-x_2)}{m^2}
\label{f2correlators} \\
\left\langle \prod_{i=1}^4 f_{\alpha_i}(x_i)
\right\rangle_{M+SD}^{U=0}   &=&  \left\langle
\prod_{i=1}^4{\tilde F}_{\alpha_i}(x_i)\cdots
\right\rangle_{M+MCS}^{U=0} + \sum_{i\ne j\ne k \ne l}
\frac{\delta_{\alpha_i\alpha_j}\delta_{\alpha_k\alpha_l}}{2m^4}
\delta (x_i-x_j)\delta (x_k-x_l)\nn \\ &-& \sum_{i\ne j\ne k \ne
l}i \frac{\delta_{\alpha_i\alpha_j}}{2m^2}\delta
(x_i-x_j)\left\langle {\tilde F}_{\alpha_k}(x_k){\tilde
F}_{\alpha_l}(x_l)\right\rangle_{M+MCS}^{U=0}
\label{f4correlators} \eea

\no Similarly, higher point correlation functions will only differ by contact terms.
The final result is such that  the map $f_{\nu}\leftrightarrow {\tilde F}_{\nu} =
\epsilon_{\nu\alpha\beta}\partial^{\alpha} A^{\beta}/m + e\, \jo /m^2 \, $ holds at
quantum level up to contact terms which vanish for non-coinciding points. Formulas
(\ref{zmsd}),(\ref{zmmcs}),(\ref{mcorrelators}),(\ref{f2correlators}) and
(\ref{f4correlators}) are strong indications that the interacting theories $\lmmcs$
and $\lmsd$  are equivalent also at quantum level. It must be stressed that the above
equivalence has been demonstrated only in the linear coupling case $\uf = 0$ which
coincides with the minimal coupling $\partial_{\nu}\to
\partial_{\nu}-\imath e f_{\nu}$ only for the case of fermions or
complex vector fields in a first order formulation as in
\cite{Fosco}. For the case of charged scalar fields the minimal
coupling prescription requires $\, U(\phi,\phi^*) = 2 e^2
\phi^*\phi \, $. Thus, in order to be able to demonstrate the
quantum equivalence we are forced to abandon the minimal coupling
for scalar fields. It is tempting to blame the second order
formulation of the scalar fields for such problems. We have tried
to rewrite them as a first order theory along the lines of
\cite{Fosco} with the introduction of auxiliary complex vector
fields. However, this has only shifted the problem from the
integral over the self-dual field to the integral over the
auxiliary vector fields which have also a badly behaved
propagator.

\section{Conclusion}

We have proposed a new master action  and proved the on shell
equivalence between the self-dual and a Maxwell Chern-Simons type
theory non-minimally coupled to $U(1)$ charged matter. The
demonstration holds for a rather general coupling to the self-dual
field and it does not depend on details of the matter sector.

At the quantum level, we have been able to prove the equivalence
between the dual models only in the linear coupling case. It is
also in this case that we have a true Maxwell Chern-Simons  theory
on the dual side. For a general coupling an awkward extra term
must be added to the MCS type theory. Although this term is on
shell ineffective it gives rise to nonrenormalizable infinities in
the quantum theory. The origin of this problem is the fact that
the self-dual field behaves like a non-dynamical field (delta
function propagator) for short distances. Usually, interaction
vertices contain only one line of a non-dynamical field ( linear
coupling ), like in the Thirring model, where the quartic
interaction can be produced by Gaussian integrating over a
non-dynamical vector field linearly coupled with the $U(1)$
fermionic current.  If we have double lines of non-dynamical
fields in the same vertex, like in
$U(\varphi,\varphi^*)f^{\nu}f_{\nu}$, the integration over such
fields will in general lead to severe infinities due to loops
containing only non-dynamical fields internal lines. In summary,
we should avoid quadratic ( and  higher ) couplings in the
self-dual field.

 Finally, the remarks made here might be useful for the
coupling with $U(1)$ matter of generalizations of the self dual
model, which has been recently investigated in
\cite{Bazeia,Menezes}, as well as for the nonabelian version of
the SD/MCS duality which is under debate in the literature
\cite{nonabelian} . The nonabelian SD model is a truly interacting
theory where the self-dual field has also a bad ultraviolet
behavior.

\section{Acknowledgements}

This work was partially supported by \textbf{CNPq}, \textbf{FAPESP} and
\textbf{CAPES}, Brazilian research agencies. We thank Alvaro de Souza Dutra  and
Marcelo Hott for discussions.

\newpage

\end{document}